\documentclass[pre,twocolumn,aps,showpacs]{revtex4-1}
\usepackage{graphicx}
\usepackage{amsmath}
\usepackage{verbatim}
\usepackage{color}
\usepackage[dvipsnames]{xcolor}
\begin{document}
\newcommand{\tca}[1]{\textcolor{blue}{#1}}
\newcommand{\tcom}[1]{{\it\textcolor{magenta}{#1}}}

\title{Lower-hybrid oscillations in a cold magnetized electron-positron-ion plasma}
\author{Prabal Singh Verma\footnote{prabal-singh.verma@univ-amu.fr}}
\affiliation{CNRS, Aix-Marseille Univ., PIIM, UMR 7345, Marseille F-13397, France}

\date{\today}

\begin{abstract}
In this paper, we obtain the dispersion relation for the lower-hybrid oscillations/waves in a cold magnetized electron-positron-ion (e-p-i) plasma. It is found that the frequency of the lower-hybrid oscillations in an e-p-i plasma is approximate $1/\sqrt{3}$ times the frequency of lower-hybrid oscillations in an electron-ion (e-i) plasma, provided the plasma density is high, and the equilibrium densities of the positron and the ion are balanced. The present work may have some relevance for laboratory/astrophysical e-p-i plasmas.
\end{abstract}


\maketitle

\section{Introduction}

Dispersion relations have always provided a fundamental base for strongly nonlinear problems in plasma physics \cite{dawson1959nonlinear, davidson1968nonlinear, davidson1972methods, stenflo1998electron, verma2010nonlinear, verma2011nonlinearkaw, brodin2014large, brodin2014nonlinear,  brodin2017simple,verma2017generation}. This is because they often allow us to gain deep insights into the experimental and astrophysical observations.  The linear dispersion relation for the electron plasma oscillations in a cold unmagnetized and magnetized plasma suggest that all the plasma species oscillate at the electron plasma frequency and the upper-hybrid frequency,
respectively \cite{chen2012introduction}. Employing a method of Lagrange variables, Davidson and Schram have examined nonlinear electrostatic oscillations for the plasmas mentioned above \cite{davidson1968nonlinear, davidson1972methods}. In both cases, the authors have obtained the exact nontrivial space-time dependent solutions in Lagrange coordinates, which depict that frequencies of the nonlinear plasma oscillations are the same as the ones derived from the corresponding linear dispersion relations. In a recent work, Brodin and Stenflo have suggested a new class of nonlinear solutions for the cold electron-plasma waves, where the physical quantities have only temporal dependence \cite{brodin2017simple}.
The authors have also obtained such solutions for
magnetized plasmas \cite{brodin2017simple}, warm plasmas \cite{stenflo2016temperature} and
dissipative plasma \cite{brodin2017nonlinear} etc.

Study of electrostatic oscillation in an e-p-i plasma has also received the recent attention of many plasma physicists, see, e.g.,
\cite{maity2014phase,verma2016nonlinear,pramanik2017effects} and references therein.
In the present work, we construct linear dispersion relations for the lower-hybrid oscillations/waves in a cold magnetized e-p-i plasma. Here we consider the dynamics of the three species (electron, positron, and ion) with and without a quasi-neutral approximation and find that the results in both cases remain the same when the plasma density is very high. Moreover, we observe that the frequency of lower-hybrid oscillations in an e-p-i plasma is rough $1/\sqrt{3}$ times the frequency of the lower-hybrid oscillations in an e-i plasma when the unperturbed densities of the positron and the ion are identical. We also demonstrate here that it is not possible to obtain the exact nonlinear solution for the lower hybrid oscillations in a cold e-p-i plasma even in the quasi-neutral approximation because the longitudinal velocities of all the species are not the same, as they are for an e-i plasma \cite{chen2012introduction,maity2010nonlinear,maity2013collisional}. Nevertheless, we believe that the linear dispersion relations as suggested here are important and should also be true for the exact nonlinear lower hybrid oscillations in a cold e-p-i plasma.

The flow of the present manuscript is organized in the following manner. In section II, equations describing the dynamics of the three plasma species (e-p-i) are introduced, and the linear dispersion relations are obtained with and without the quasi-neutral criterion.
Section III contains the summary of the results and discussion.

\section{Governing equations}
The basic equations describing the dynamics of a cold magnetized e-p-i plasma are the continuity equations,
\begin{equation}\label{econ}
 {\partial_t n_{e}}+\boldsymbol{\nabla}\cdot(n_{e}{\bf v_{e}})=0,
\end{equation}
\begin{equation}\label{pcon}
 {\partial_t n_{p}}+\boldsymbol{\nabla}\cdot(n_{p}{\bf v_{p}})=0,
\end{equation}
\begin{equation}\label{icon}
 {\partial_t n_{i}}+\boldsymbol{\nabla}\cdot(n_{i}{\bf v_{i}})=0,
\end{equation}
{the}
momentum equations,
\begin{equation}\label{emom}
({\partial_t } + {\bf v_e} \cdot \boldsymbol{\nabla}){\bf v_e}= \frac{q_e}{m_e} [{\bf E} + (1/c){\bf v_e}\times {\bf B}],
\end{equation}
\begin{equation}\label{pmom}
({\partial_t } + {\bf v_p} \cdot \boldsymbol{\nabla}){\bf v_p}= \frac{q_p}{m_p} [{\bf E} + (1/c){\bf v_p}\times {\bf B}],
\end{equation}
\begin{equation}\label{imom}
({\partial_t } + {\bf v_i} \cdot \boldsymbol{\nabla}){\bf v_i}= \frac{q_i}{m_i} [{\bf E} + (1/c){\bf v_i}\times {\bf B}],
\end{equation}
Poisson's equation,
 \begin{equation}\label{poi}
   \boldsymbol{\nabla} \cdot {\bf E}= 4 \pi \sum_{\alpha} q_{\alpha} n_{\alpha}%e (n_p - n_e).
 \end{equation}
and Ampere's Law,
 \begin{equation}\label{curr}
   \boldsymbol{\nabla} \times {\bf B}= \frac{4 \pi}{c}  {\bf J} + \frac{4 \pi}{c} {\partial_t {\bf E}}
 \end{equation}

Here the subscript `$e$,' stands for the electron, `$p$,' for the
positron and `$i$,' for the ion. The densities, velocities, masses, and charges of all the three species are denoted by
$n_\alpha$, $\mathbf{v_\alpha}$, $m_\alpha$ and $q_\alpha$, respectively, where $\alpha$ is the corresponding subscript. The quantity $c$ is the speed of light in vacuum, ${\bf E}$ the electric field and ${\bf B}$ is a homogeneous external magnetic field applied along the 
$z$-direction, i.e. ${\bf B} = B_0 \hat z$.
{\it CGS} unit is used throughout. Spatial variations are restricted to one direction, which we consider to be along the $x$-axis,
without any loss of generality.  Thus, the set of equations
\eqref{econ}-\eqref{curr} reduces to the following equations, 

\begin{equation}\label{econ1}
 {\partial_t n_{e}}+{\partial_x(n_{e}v_{ex})}=0,
\end{equation}
\begin{equation}\label{pcon1}
 {\partial_t n_{p}}+{\partial_x(n_{p}v_{px})}=0,
\end{equation}
\begin{equation}\label{icon1}
 {\partial_t n_{i}}+{\partial_x(n_{i}v_{ix})}=0,
\end{equation}
\begin{equation}\label{emom1_x}
{\partial_t v_{ex}} + v_{ex}{\partial_x v_{ex}}= -\frac{q_e}{m_e} \partial_x \phi - \omega_{ce} v_{ey},  
\end{equation}
\begin{equation}\label{pmom1_x}
{\partial_t v_{px}} + v_{px}{\partial_x v_{px}}=  -\frac{q_p}{m_p} \partial_x \phi + \omega_{cp} v_{py}, 
\end{equation}
\begin{equation}\label{imom1_x}
{\partial_t v_{ix}} + v_{ix}{\partial_x v_{ix}}=  -\frac{q_i}{m_i} \partial_x \phi + \omega_{ci} v_{iy}, 
\end{equation}
\begin{equation}\label{emom1_y}
{\partial_t v_{ey}} + v_{ex}{\partial_x v_{ey}}=   \omega_{ce} v_{ex},  
\end{equation}
\begin{equation}\label{pmom1_y}
{\partial_t v_{py}} + v_{px}{\partial_x v_{py}}=  - \omega_{cp} v_{px}, 
\end{equation}
\begin{equation}\label{imom1_y}
{\partial_t v_{iy}} + v_{ix}{\partial_x v_{iy}}=  - \omega_{ci} v_{ix}, 
\end{equation}
  \begin{equation}\label{poi1}
  -{\partial_{xx} \phi}= 4 \pi e (n_i + n_p - n_e),
\end{equation}
\begin{equation}\label{curr1}
 0 = 4 \pi J_x + {\partial_t E_x} 
\end{equation}

where $\omega_{c\alpha} = |q_{\alpha} B_0/m_{\alpha}c|$ denotes the cyclotron frequency of the plasma species. Here we have expressed electric field $E_x$, as the gradient of the electric potential $\phi$, i.e., $E_x = -\partial_x \phi $. We note here that in Eq.\eqref{curr1} L.H.S. is zero because the magnetic field is constant.

Now we proceed to adopt the linear analysis by ignoring the nonlinear terms in the above equations. For instance, we 
can write down Eqs.\eqref{econ1}-\eqref{poi1} in the linearized form as follows,
\begin{equation}\label{econ2}
 {\partial_t n_{e}^{(1)}}+n_{0e}{\partial_x v_{ex}^{(1)}}=0,
\end{equation}
\begin{equation}\label{pcon2}
 {\partial_t n_{p}^{(1)}}+n_{0p}{\partial_x v_{px}^{(1)}}=0,
\end{equation}
\begin{equation}\label{icon2}
 {\partial_t n_{i}^{(1)}}+n_{0i}{\partial_x v_{ix}^{(1)}}=0,
\end{equation}
\begin{equation}\label{emom2_x}
{\partial_t v_{ex}^{(1)}} = \frac{e}{m_e} \partial_x \phi^{(1)} - \omega_{ce} v^{(1)}_{ey},  
\end{equation}
\begin{equation}\label{pmom2_x}
{\partial_t v_{px}^{(1)}} =  -\frac{e}{m_p} \partial_x \phi^{(1)} + \omega_{cp} v^{(1)}_{py}, 
\end{equation}
\begin{equation}\label{imom2_x}
{\partial_t v_{ix}^{(1)}} =  -\frac{e}{m_i} \partial_x \phi^{(1)} + \omega_{ci} v^{(1)}_{iy}, 
\end{equation}
\begin{equation}\label{emom2_y}
{\partial_t v^{(1)}_{ey}} =   \omega_{ce} v_{ex}^{(1)},  
\end{equation}
\begin{equation}\label{pmom2_y}
{\partial_t v^{(1)}_{py}} =  - \omega_{cp} v_{px}^{(1)}, 
\end{equation}
\begin{equation}\label{imom2_y}
{\partial_t v^{(1)}_{iy}} =  - \omega_{ci} v_{ix}^{(1)}, 
\end{equation}
  \begin{equation}\label{poi2}
  -{\partial_{xx} \phi^{(1)}}= 4 \pi e \big[n_i^{(1)} + n_p^{(1)} - n_e^{(1)}\big],
\end{equation}

where the superscript $(1)$ represents the linear approximation and we assume $q_i = q_p = e$, $q_e = -e$, considering ions to be singly ionized. The quantities $n_{0e}, n_{0p}$ and $n_{0i}$ denote the equilibrium densities of electron, positron and ion, respectively. They satisfy the following criterion due to the charge neutrality,
\begin{equation}\label{equ_den}
n_{0e} = n_{0p} + n_{0i}
\end{equation}

\subsection{Standard dispersion relation}
In order to obtain the dispersion relation for the lower-hybrid oscillations/waves, we assume that all the physical quantities have the following form,
\begin{equation}\label{fourier}
f(x,t) \propto  exp[i(kx-\omega t)], 
\end{equation}
where $k$ and $\omega$ are the wave number and the wave frequency, respectively. Thus, the partial derivatives $\partial_t$ and $\partial_x$ in Eqs.\eqref{econ2}-\eqref{poi2} can be
replaced with $-i\omega$ and $ik$, respectively. After some algebra we arrive at the following relations,

\begin{eqnarray} 
n_e^{(1)} &=& \frac{k}{\omega} n_{0e} v_{ex}^{(1)}, \label{ne1}\\
n_p^{(1)} &=& \frac{k}{\omega} n_{0p} v_{px}^{(1)}, \label{np1}\\
n_i^{(1)} &=& \frac{k}{\omega} n_{0i} v_{ix}^{(1)}, \label{ni1}\\
v_{ex}^{(1)} &=& -\frac{ek}{m\omega}\Big[ 1 - \frac{\omega_{c}^2}{\omega^2}\Big]^{-1} \phi^{(1)}, \label{ve1}\\ 
v_{px}^{(1)} &=& \frac{ek}{m\omega}\Big[ 1 - \frac{\omega_{c}^2}{\omega^2}\Big]^{-1} \phi^{(1)}, \label{vp1}\\ 
v_{ix}^{(1)} &=& \frac{ek}{m_i\omega}\Big[ 1 - \frac{\omega_{ci}^2}{\omega^2}\Big]^{-1} \phi^{(1)}, \label{vi1}\\
k^2 \phi^{(1)} &=& 4 \pi e \big[n_i^{(1)} + n_p^{(1)} - n_e^{(1)}\big],  
\end{eqnarray}
Here we have assumed $m_e = m_p = m$ and $\omega_{ce} = \omega_{cp}=\omega_{c}$.
Above equations can be further combined to give the following,
\begin{eqnarray} \label{dispersion1}
1 - \frac{\omega_{pi}^2}{\omega^2-\omega_{ci}^2} - \frac{\omega_{pe}^2}{\omega^2-\omega_{c}^2} - \frac{\omega_{pp}^2}{\omega^2-\omega_{c}^2} &=& 0,  
\end{eqnarray}
where $\omega_{pi}$, $\omega_{pe}$ and $\omega_{pp}$ are the plasma frequencies of the ion, electron and positron,
respectively and their expressions are,

 \begin{eqnarray} 
\nonumber \omega_{pi} = \sqrt{\frac{4 \pi n_{0i}e^2}{m_i}}, \omega_{pe} = \sqrt{\frac{4 \pi n_{0e}e^2}{m}}, \omega_{pp} = \sqrt{\frac{4 \pi n_{0p}e^2}{m}}, \\&
\end{eqnarray}
Since we are here interested in the wave with frequency $\omega > \omega_{ci}$, we can drop $\omega_{ci}^2$ from
Eq.\eqref{dispersion1} and obtain the relation,
\begin{eqnarray} \label{dispersion2}
\omega^4 - (\omega_{c}^2+\omega_{pi}^2+\omega_{pp}^2+\omega_{pe}^2)\omega^2 + \omega_{pi}^2 \omega_{c}^2 &=& 0,  
\end{eqnarray}
We can further drop the term $\omega_{pi}^2$ being much smaller than the other terms, e.g. $\omega_{pe}^2, \omega_{c}^2$, etc,
and the Eq.\eqref{dispersion2} takes the form,
\begin{eqnarray} \label{dispersion3}
\omega^4 - (\omega_{c}^2+\omega_{pp}^2+\omega_{pe}^2)\omega^2 + \omega_{pi}^2 \omega_{c}^2 &=& 0,  
\end{eqnarray}
Eq.\eqref{dispersion3} is quadratic in $\omega^2$ and the solution for the same can be written as,
\begin{eqnarray} \label{dispersion4}
\nonumber \omega^2 &=& \frac{(\omega_{c}^2+\omega_{pp}^2+\omega_{pe}^2) \pm \sqrt{(\omega_{c}^2+\omega_{pp}^2+\omega_{pe}^2)^2-4\omega_{pi}^2 \omega_{c}^2}}{2}, \\ & 
\end{eqnarray}
where the `$+$' sign corresponds to the upper-hybrid oscillations and the `$-$' sign stands for the lower-hybrid oscillations.
Since our interest lies in the lower-hybrid waves, we consider the `$-$' sign and get the following,
\begin{eqnarray} \label{dispersion5}
\nonumber \omega^2 &=& \frac{1}{2}\Bigg[(\omega_{c}^2+\omega_{pp}^2+\omega_{pe}^2) \\ \nonumber && - (\omega_{c}^2+\omega_{pp}^2+\omega_{pe}^2) 
\sqrt{1-\frac{4\omega_{pi}^2 \omega_{c}^2}{(\omega_{c}^2+\omega_{pp}^2+\omega_{pe}^2)^2}}\Bigg], \\ & 
\end{eqnarray}
however,
 \begin{eqnarray} %\label{dispersion5}
\nonumber \frac{4\omega_{pi}^2 \omega_{c}^2}{(\omega_{c}^2+\omega_{pp}^2+\omega_{pe}^2)^2} \ll 1. 
\end{eqnarray}
This allows us to employ Taylor expansion for the term in the square root of the Eq.\eqref{dispersion5}. Ignoring smaller
terms we get the relation,
 \begin{eqnarray} \label{dispersion6}
 \frac{1}{\omega^2} &=& \frac{1}{\omega_{pi}^2} + \frac{1+2n_{0p}/n_{0i}}{ \omega_{c} \omega_{ci}}, %\\ & 
\end{eqnarray}
where $\omega$ in Eq.\eqref{dispersion6} is the frequency of the lower-hybrid oscillations in a cold magnetized e-p-i plasma. Note here that the first term on the R.H.S. of the Eq.\eqref{dispersion6} dominates for low-density plasmas and cannot be ignored. Nevertheless, for high-density plasmas, it becomes much smaller as compared to the second term and hence can be dropped safely for such plasmas. Thus, for high density cold e-p-i plasmas we arrive at the following expression for the lower-hybrid oscillations,
 \begin{eqnarray} \label{dispersion7}
 {\omega} &\approx&  \sqrt{\frac{ \omega_{c} \omega_{ci}}{1+2n_{0p}/n_{0i}}}, %\\ & 
\end{eqnarray}
We show in the next subsection that above expression for the lower-hybrid oscillations results naturally in the quasi-neutral approximation. However, before proceeding to do that, we note
here that as the density of the positron $n_{0p}\rightarrow 0$, we recover earlier results for e-i plasmas \cite{chen2012introduction,maity2010nonlinear}.

\subsection{Dispersion relation with quasi-neutrality}
In the quasi-neutrality, instead of solving the Poisson's equation we employ the plasma approximation,
\begin{equation}\label{quasi}
n_i +n_p = n_e, 
\end{equation}
Moreover, the displacement current term ($\partial_t E_x $)
can also be safely dropped from Eq.\eqref{curr1} because here we are interested in study of low frequency oscillations.
This implies that total current density along the $x$-direction is zero, i.e.
\begin{equation}\label{curr2}
J_x = (q_i n_i v_{ix} +q_p n_p v_{px}+q_e n_e v_{ex}) = 0. 
\end{equation}
Thus, Eq.\eqref{curr2} becomes,
\begin{equation}\label{curr12}
n_i v_{ix} +n_p v_{px} = n_e v_{ex}.
\end{equation}
Although from Eqs.\eqref{quasi}-\eqref{curr2} it may seem that the longitudinal velocities of all the three species $v_{\alpha x}$ are equal, but we will see shortly that it cannot be true. Now writing Eqs.\eqref{quasi}-\eqref{curr2} in the first order approximation,

\begin{equation}\label{quasi1}
n_i^{(1)} +n_p^{(1)} = n_e^{(1)}, 
\end{equation}
\begin{equation}\label{curr3}
n_{0i} v_{ix}^{(1)} +n_{0p} v_{px}^{(1)} = n_{0e} v_{ex}^{(1)}.
\end{equation}
We now use Eqs.\eqref{ne1}-\eqref{vi1} in Eq.\eqref{quasi1} to obtain,
\begin{eqnarray} \label{dispersion8}
\frac{\omega_{pi}^2}{\omega^2-\omega_{ci}^2} + \frac{\omega_{pe}^2}{\omega^2-\omega_{c}^2} + \frac{\omega_{pp}^2}{\omega^2-\omega_{c}^2} &=& 0,  
\end{eqnarray}
Above equation after some algebra gives the following dispersion relation,
 \begin{eqnarray} \label{dispersion9}
 {\omega} &\approx&  \sqrt{\frac{ \omega_{c} \omega_{ci}}{1+2n_{0p}/n_{0i}}}, %\\ & 
\end{eqnarray}
here we have used the fact that $\omega^2 \gg \omega_{ci}^2$. We thus find that Eq.\eqref{dispersion7} and Eq.\eqref{dispersion9} are identical, and this indicates that the quasi-neutrality is a reasonable approach to study lower-hybrid oscillations in a cold magnetized e-p-i plasma provided the
plasma density is very high. We also learn here that when the densities of the positron and the ion are equal, we have ${\omega} \approx  \sqrt{\omega_{c}\omega_{ci}/3}$. Therefore, for balanced positron and ion densities, the frequency of lower-hybrid oscillations in an e-p-i plasma is approximate $1/\sqrt{3}$ times the frequency of lower-hybrid oscillations in an e-i plasma.

We further note here that one would get the same expression for the lower-hybrid frequency when the linearized current 
equation \eqref{curr3} is used. This gives a clue that longitudinal velocities of the three species cannot be the same, because if we just employ $v_{ex}^{(1)} = v_{ex}^{(1)} =v_{ix}^{(1)}$, the expression for the lower-hybrid frequency are not recovered. 
We thus conclude that the method of Lagrange variables cannot be used here to obtain an exact solution for the nonlinear lower-hybrid oscillations in a cold magnetized e-p-i plasma because the Lagrange technique is applicable only when all the plasma species have the 
same longitudinal velocity, as in e-i plasmas under quasi-neutral assumption \cite{maity2010nonlinear}.
%We conclude that the exact nonlinear solutions are not possible here because different plasma species have different
%longitudinal velocities.

{\section{Summary and discussion}}
In summary, we have obtained linear dispersion relations for the lower-hybrid oscillations in a cold magnetized e-p-i plasma with and without {the quasi-neutral approximation}. We have found that the quasi-neutrality is a valid approximation for high-density e-p-i plasmas. Moreover, we have shown that for equal densities of the ion and the positron, the frequency of the lower-hybrid oscillations in an e-p-i plasma is approximate $1/\sqrt{3}$ times the frequency of lower-hybrid oscillations in an e-i plasma.
We have further demonstrated that exact solutions for the lower hybrid oscillations in a cold e-p-i plasma are not possible due to different longitudinal velocities of different species. We conclude by saying that although the results in the present paper are linear, they might be 
relevant to the laboratory and astrophysical e-p-i plasmas because we believe that the corresponding nonlinear lower-hybrid oscillations/waves will obey the same frequency as suggested here. We will verify this statement shortly by employing our recently developed code \cite{verma2018novel} to study these oscillations numerically. 

\vspace{1cm}
\section*{Acknowledgments} {The author would like to express his sincere thanks to the anonymous referee for his/her critical and motivating remarks which has given an entirely new shape to the present work.} Moreover, useful discussions with Dr. Yannick Marandet are also greatly acknowledged. This work is funded by CNRS and Aix-Marseille University, France.

%\bibliography{lib_plasma_osci}
%

%\bibitem [{\citenamefont {Verma}(2018)}]{verma2018novel}%
%  \BibitemOpen
%  \bibfield  {author} {\bibinfo {author} {\bibfnamefont {P.~S.}\ \bibnamefont
%  {Verma}},\ }\href@noop {} {\bibfield  {journal} {\bibinfo  {journal} {Physics of Plasmas, In Press}\ } (\bibinfo {year} {2018})}\BibitemShut {NoStop}%

\end{document}